\documentclass[a4paper, twocolumn,showpacs,preprintnumbers,amsmath,amssymb, superscriptaddress, prl, floatfix]{revtex4}
\usepackage{graphicx}
\usepackage{dcolumn}
\usepackage{bm}

\usepackage{color,ulem}

\begin{document}
\title{Spatial Modulation Microscopy for Real-Time Imaging of Plasmonic Nanoparticles and Cells}
\author{N. Fairbairn} \affiliation{Institute for Life Sciences and Faculty of Physical and Applied Sciences, University of Southampton, Highfield, Southampton
SO17, 1BJ, United Kingdom}
\author{R. A. Light}
\affiliation{Applied Optics Group, Faculty of Engineering,
University of Nottingham, Nottingham NG7 2RD, United Kingdom.}
\author{R. Carter}\affiliation{Institute for Life Sciences and Faculty of
Medicine, University of Southampton, United Kingdom}
\author{R. Fernandes}\affiliation{Institute for Life Sciences and Faculty of Physical and Applied Sciences, University of Southampton, Highfield, Southampton
SO17, 1BJ, United Kingdom}
\author{A. G. Kanaras} \affiliation{Institute for Life Sciences and Faculty of Physical and Applied Sciences, University of Southampton, Highfield, Southampton
SO17, 1BJ, United Kingdom}
\author{T. J. Elliott}
\affiliation{Institute for Life Sciences and Faculty of
Medicine, University of Southampton, United Kingdom}
\author{M. G. Somekh}
\affiliation{Applied Optics Group, Faculty of Engineering,
University of Nottingham, Nottingham NG7 2RD, United Kingdom.}
\author{M. C. Pitter}
\affiliation{Applied Optics Group, Faculty of Engineering,
University of Nottingham, Nottingham NG7 2RD, United Kingdom.}
\author{O. L. Muskens}
\affiliation{Institute for Life Sciences and Faculty of Physical and Applied Sciences, University of Southampton, Highfield, Southampton
SO17, 1BJ, United Kingdom}\email{O.Muskens@soton.ac.uk}

\date{\today}

\begin{abstract}
Spatial modulation microscopy is a technique originally developed
for quantitative spectroscopy of individual nano-objects. Here, a
parallel implementation of the spatial modulation microscopy
technique is demonstrated based on a line detector capable of
demodulation at kHz frequencies. The capabilities of the imaging
system are shown using an array of plasmonic nanoantennas and
dendritic cells incubated with gold nanoparticles.
\end{abstract}

\maketitle

With the increasing employment of nanomaterials in physical and
biomedical science, sensitive methods capable of screening and
characterizing these materials are needed. Currently there is a
range of optical microscopy techniques capable of imaging
individual nanoparticles \cite{vandijk}. For nonfluorescent
particles, interactions take place through absorption and/or
scattering. Detection of scattered intensity, generally known as
darkfield microscopy, is the most widely used technique, combining
background-free imaging with a high spectral selectivity for e.g.
optical sensing \cite{molecularruler, duyne}. The efficiency of
light scattering is greatly reduced with decreasing particle size,
proportional to the square of the particle volume, which can be
overcome using more advanced interferometric detection schemes
\cite{vandijk,kukuraiscat}. Also scattering-based techniques are
not suitable for materials with an a small dielectric contrast,
i.e. index-matched particles in solution.

A fundamentally different method of detection relies on detection
of absorption rather than scattering of light. The most sensitive
of these techniques is photothermal imaging, where a thermal
response caused by local absorption can be detected with
single-molecule sensitivity \cite{orritscience2010}. Photothermal
imaging methods require a relatively high laser intensity
(MW/cm$^2$) over a diffraction limited spot, which limits the
scalability of photothermal imaging in real-time and live cell
applications. Spatial modulation microscopy (SMM) has been
introduced as a technique for quantitative analysis of the optical
extinction cross-section of small metal nanoparticles
\cite{arbouet2004smm, muskensjpc2008}. The method is based on recovery of a small
modulation component in the transmitted (or reflected) light using
lock-in detection. This spatial modulation is achieved by means of
a periodical displacement of the specimen in a Gaussian laser
focus. While the technique is less sensitive than photothermal
imaging, it allows for precise in-situ quantitative spectroscopy
of nanomaterials which can then be correlated to other properties
such as their ultrafast response \cite{martinaantennas}. In
addition, the SMM technique is scalable as it works at moderate optical
intensities, i.e. a few W/cm$^2$ per pixel, and requires only a small
focal width in one dimension. No reports have been made yet on the
potential of SMM for real-time imaging in biological systems.

\begin{figure}[b]
    \centering
        \includegraphics[width=8.2cm]{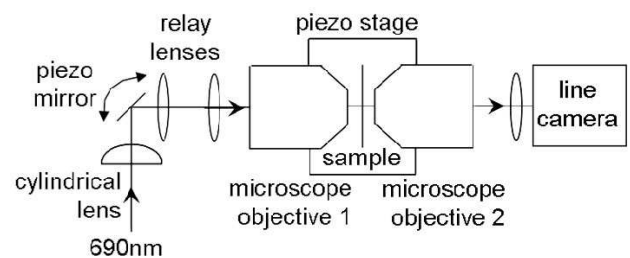}
    \caption{Experimental setup for spatial modulation microscopy.}
    \label{fig:smmsetup}
\end{figure}

Here, we demonstrate the integration of SMM into a viable imaging
system covering an area of tens of micrometers at a rate of around
one image per second. The fast scanning spatial modulation method,
makes use of a recently developed CMOS camera technology combining
multiple wells and a phase stepping technique to detect a
modulating signal \cite{nottcamera1,nottcamera2}. The line camera
replaces conventional lock-in amplification to detect the
modulating signal produced by the spatial modulation technique.
The camera features a pixel well depth of 2 billion electrons, providing
a theoretical shot noise level of $2\times 10^{-5}$
per well. In addition, a series of four on-chip wells per pixel
allows fast acquisition of four frames in each modulation period,
which can be used to recover the first and second harmonic spatial modulation at kHz frequencies. The resulting intensities are converted into 16-bit integers, each count representing $3\times10^4$ electrons.

\begin{figure}[t]
    \centering
        \includegraphics[width=7.3cm]{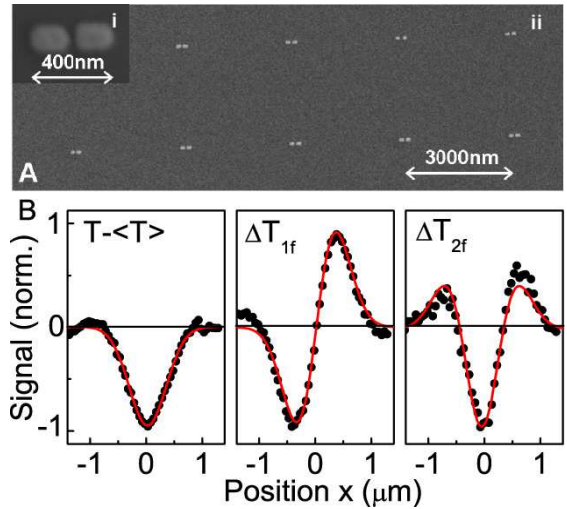}
    \caption{a) SEM images of nanoantennnas, with close up of single antenna. (b) Experimental data for a single antenna for intensity $T-\left<T\right>$, and SMM profiles at first ($\Delta T_{1f}$) and second harmonic ($\Delta T_{2f}$) of the modulation frequency, together with fits (lines) to the derivatives of a Gaussian beam profile of $0.8\pm 0.01$~$\mu$m width.}
    \label{fig:smmfits}
\end{figure}

The experimental setup is shown in Fig.~\ref{fig:smmsetup}. A
transmission microscope was used consisting of two $100 \times$,
0.9 NA microscope objectives. A vertical line focus was created
from a 690~nm beam by imaging of the focus of a cylindrical lens
onto the sample. The transmitted light was subsequently collected
by the second objective and imaged onto the CMOS line array.
Spatial modulation was implemented by periodically tilting of a
custom-built piezo-actuated flexure mirror at a frequency of
1.3~kHz. The microscopy objective acts as a fourier element
converting the angular tilt into a spatial displacement of the
line focus. This method greatly improves the flexibility of the
SMM method in combination with bulky or liquid specimens compared
to previous implementations based on a piezoelectric sample
displacement \cite{arbouet2004smm}. In order to
optimally exploit the dynamic range of the 256-pixel array at a
typical frame rate of 5200 frames per second, a total illumination
power of around 2~mW is required. Given the optical resolution of
0.2 $\mu$m per pixel this results in an average local intensity of
around $10$~W/cm$^2$.

\begin{figure}[t]
    \centering
        \includegraphics[width=8.5cm]{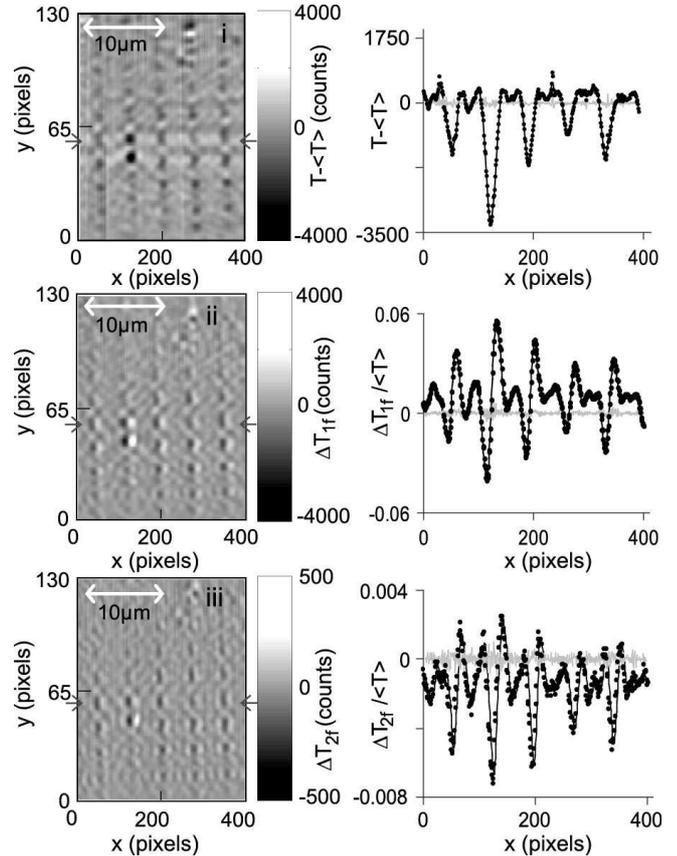}
    \caption{Spatial modulation microscopy maps of intensity difference from average (i), first harmonic (ii) and second
    harmonic components
    (iii) of the spatial modulation signal. Line graphs (I-III) are cross sections of the maps normalized to the average intensity, with (grey lines) noise background after subtraction of moving average.}
    \label{fig:smmantennas}
\end{figure}

The SMM imaging performance was tested by mapping an array of
lithographically defined gold nanoantennas on an ITO coated glass
substrate. The nanoantennas, shown in
Fig.~\ref{fig:smmfits}(a), comprise of two arms of 200~nm in
length and 100~nm wide, with a a thickness of 25~nm, and were
designed to have a resonant dipole mode at around 850~nm
\cite{martinaantennas}. Two-dimensional SMM maps were made of the transmitted intensity $T$ as well as
the first and second harmonic components of the spatial modulation
intensity, respectively denoted as $\Delta T_{1f}$ and $\Delta T_{2f}$
\cite{arbouet2004smm}. For the intensity graph, an average line
frame in absence of antennas (denoted as $\left<T\right>$) was subtracted to
remove a constant variation in pixel illumination. Figure~\ref{fig:smmfits}(b) shows typical cross-sections through a single antenna selected from a larger image shown in Fig.~\ref{fig:smmantennas}. The signals can be well fitted to a Gaussian beam profile for $T$ and its first and second derivatives for $\Delta T_{1f}$ and $\Delta T_{2f}$, as expected from theory \cite{arbouet2004smm}. The fits yield a full-width-at-half-maximum of $0.8\pm 0.1$~$\mu$m. This is larger than the diffraction limit due to aberrations in the imaging system. The depth of focus of the SMM signals follows the divergence of the Gaussian beam and results in a decay over $\sim 1.5$~$\mu$m. The absence of an out-of-focus background signal is of potential use in imaging of three-dimensional specimens.

\begin{figure}[htb]
    \centering
        \includegraphics[width=8.5cm]{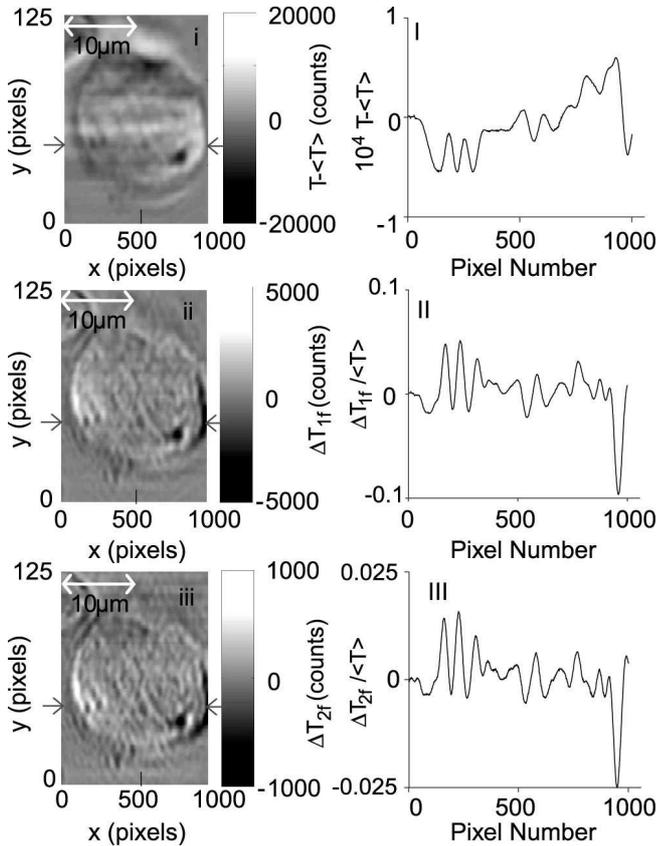}
    \caption{i - Intensity, ii - 1f and iii - 2f SMM images of dendritic cells along with a cross section.}
    \label{fig:smmcellscrosssection}
\end{figure}

Figure \ref{fig:smmantennas} shows the SMM image obtained for a
$20\times 25$~$\mu$m$^2$ area containing nanoantennas. Each line
was averaged over 4 frames, resulting in a total acquisition time
of 2~seconds. This is an improvement in collection time by more
than two orders of magnitude compared to conventional
point-scanning SMM \cite{arbouet2004smm}. Only a central 120-pixel
part of the 256-pixel array in the y-direction is displayed
containing the antenna array. Horizontal cross-sections through a
row of antennas (indicated by arrows) are shown in
Fig.~\ref{fig:smmantennas}(b), where the $1f$ and $2f$ components
have been normalized to the average intensity $\left<T\right>$ to
obtain the relative spatially modulated transmission.

The noise level of the measurement was calculated by subtracting a 5 point moving
average from the original traces taken at a horizontal step size
of 50~nm. Resulting noise traces are given by the grey lines in
the cross sections of Fig.~\ref{fig:smmantennas}(b). We obtained
average noise levels of $3\times 10^{-4}$ and $1\times 10^{-4}$
for the $1f$ and $2f$ signals respectively. The noise level is
exceeds the shot-noise due to the presence of some technical noise in our system at kHz frequencies. The signal to noise ratio exceeds 100 for the antennas under study at the first
harmonic frequency, demonstrating the capability of the
system to image individual nanostructures.

In a second experiment, a sample of dendritic cells was incubated
with 5$\mu$l of a CTAB-coated gold nanorod solution for 2 hours. The
gold nanorods, of $16\times42$~nm$^{2}$ average dimensions, have a
strong optical resonance at around 690~nm corresponding to a
longitudinal surface plasmon resonance. Figure
\ref{fig:smmcellscrosssection} shows an image of a dendritic cell
obtained using the intensity, $1f$ and $2f$ components of the SMM signal, along with
a cross section of each image taken at the position indicated by the arrows.
The expected SMM signal from individual nanorods is around
$10^{-3}$ \cite{muskensjpc2008}, i.e. much smaller than the cellular background signals in Fig.~\ref{fig:smmcellscrosssection}. The lack of specific contrast thus limits the efficacy of
SMM in imaging nanoparticles in cells compared other methods such
as photothermal imaging. The $1f$ and $2f$ components give the first and second
derivatives of the light transmitted through the cell, which results in an enhancement of edges
and point-like objects. This sensitivity of the SMM
signal to small variations in optical density in the cell may be of interest as an alternative to phase-sensitive imaging, i.e. phase-contrast or Nomarski, under conditions that phase information is distorted, e.g. for very thick or turbid
samples.

In conclusion, a parallel implementation of spatial modulation
 microscopy has been demonstrated. The theoretical sensitivity of the CMOS demodulation is comparable to state-of-the-art lock-in amplifiers, and the current system allows  resolving an array of plasmonic nanoantennas of 200~nm with a signal-to-noise ratio of 100. This may be improved by
reducing sources of classical noise in the system, optimization of the diffraction-limited focus, and averaging over pixels and/or
time. Key aspects of SMM are retained by the setup while improving the acquisition time by several orders of
magnitude, therefore opening up new applications requiring
simultaneous imaging and measurement of multiple nanoparticles
e.g. in solid or microfluidic environments. The absence of out-of-focus signals and availability of higher-order derivatives make SMM a potentially useful alternative or addition to phase-contrast techniques.

The authors acknowledge M. Abb for fabrication of nanoantennas and
H. Warren for help with cell preparation.

\end{document}